\documentclass[preprint,aps,prd,amsmath,amssymb,nofootinbib]{revtex4}
\usepackage{graphicx}
\usepackage{color}
\usepackage[latin1]{inputenc}
\usepackage{ulem}
\newcommand{\be}{\begin{eqnarray}}
\newcommand{\beq}{\begin{equation}}
\newcommand{\eeq}{\end{equation}}
\newcommand{\ee}{\end{eqnarray}}

\newcommand{\bmp}{\noindent\begin{minipage}{16cm}}
\newcommand{\emp}{\end{minipage}\vskip 7mm} 

\usepackage{graphicx}
\usepackage{dcolumn}
\usepackage{bm}
\usepackage{amsmath}
\usepackage{amsfonts}
\usepackage{bbm}
\usepackage{subfigure}

\def\drawbox#1#2{\hrule height#2pt
        \hbox{\vrule width#2pt height#1pt \kern#1pt
              \vrule width#2pt}
              \hrule height#2pt}

\def\Asym#1#2{\vcenter{\vbox{\drawbox{#1}{#2}
              \kern-#2pt 
              \drawbox{#1}{#2}}}}

\def\({\left(}
\def\){\right)}

\begin{document}

\title{Can Neutron stars constrain Dark Matter?}
\author{Chris {\sc Kouvaris}}\email{ckouvari@ulb.ac.be}
\author{Peter {\sc Tinyakov}}\email{Petr.Tiniakov@ulb.ac.be}
 \affiliation{Service de Physique Th\'eorique,  Universit\'e Libre de Bruxelles, 1050 Brussels, Belgium}

\begin{abstract}
We argue that observations of old neutron stars can impose constraints
on dark matter candidates even with very small elastic or inelastic
cross section, and self-annihilation cross section. We find that old
neutron stars close to the galactic center or in globular clusters can
maintain a surface temperature that could in principle be
detected. Due to their compactness, neutron stars can accrete WIMPs
efficiently even if the WIMP-to-nucleon cross section obeys the
current limits from direct dark matter searches, and therefore they
could constrain a wide range of dark matter candidates.
\end{abstract}


\maketitle
\section{Introduction}

Since the initial discovery of the ``missing mass'' problem by Zwicky in
the 30's, a lot of theoretical, experimental, and observational effort
has been put in unveiling the mystery of dark matter. A number of
possibilities have been proposed, including modifications of the
gravitational theory, hidden sector(s), primordial black holes and other
massive objects, and new dark matter particles.

An attractive solution of the dark matter problem within the context
of particle physics can be provided by a class of models with Weakly
Interacting Massive Particles (WIMP). The Standard Model does not have
a WIMP with the required characteristics, which means that WIMPs are
probably related to physics beyond the Standard Model. There are
several dark matter propositions according to what extension of the
Standard Model one selects: supersymmetry
\cite{Jungman:1995df,Bertone:2004pz}, hidden
sectors~\cite{Pospelov:2007mp,Hambye:2008bq}, technicolor
~\cite{Gudnason:2006yj,Kouvaris:2007iq,Ryttov:2008xe,Sannino:2010ia},
etc.

All currently existing evidence in favor of dark matter (as, for
example, WMAP~\cite{Dunkley:2008ie}) is of gravitational origin.  In
order to distinguish between the dark matter models, a direct
(non-gravitational) detection of dark matter particles is
required. The most important parameters that determine the
perspectives of the direct detection are the cross section $\sigma_N$
of the dark matter-to-nucleon interaction, and the dark matter
self-annihilation cross section $\sigma_A$, or the decay rate in
models with decaying dark matter. Underground direct search
experiments such as CDMS~\cite{Ahmed:2009zw} and
Xenon~\cite{Angle:2008we} have put tight constraints on the
spin-independent and spin-dependent cross sections of WIMPs scattering
off nuclei targets at the level of $\sigma_N\lesssim 10^{-43}{\rm
  cm}^2$. Interestingly, the DAMA collaboration~\cite{Bernabei:2010mq}
claims the observation of an annual modulated signal with high
statistical significance.  A possible reconciliation of all the
underground search experiments points to the existence of dark matter
with excited states, in which case WIMPs can interact also
inelastically~\cite{TuckerSmith:2001hy,TuckerSmith:2004jv}, or to less
mainstream scenarios as in~\cite{Khlopov:2007ic,Khlopov:2008ty}.

In the last twenty years, there have been several attempts to
constrain the properties of WIMPs by looking at signatures related to
the accretion and/or annihilation of WIMPs inside stars. This includes
the capture of WIMPs in the Earth and the
Sun~\cite{Press:1985ug,Gould:1987ju,Gould:1987ww}, the
self-annihilation of WIMPs that can lead to an observable neutrino
spectrum~\cite{Jungman:1994jr,Nussinov:2009ft}, the effect of dark
matter in the evolution of low-mass
stars~\cite{Casanellas:2009dp,Casanellas:2010sj}, and the study of the
WIMP accretion and/or annihilation inside compact stars such as
neutron stars ~\cite{Goldman:1989nd,Kouvaris:2007ay,Sandin:2008db} and
white dwarfs~\cite{Bertone:2007ae,McCullough:2010ai}.

Compact objects, and in particular, neutron stars constitute a
potentially promising way of constraining dark matter models. Firstly,
the high baryonic density in compact stars increases the probability
of WIMP scattering within the star and eventually the gravitational
trapping. This is crucial in view of the tiny value of $\sigma_N$. It
should be noted that in the models with the inelastic dark matter
interactions, the elastic and inelastic cross sections of the WIMP
scattering inside the star are of the same order, because the WIMP
velocity is much higher that the asymptotic value of
$220 \text{km/s}$, and its kinetic energy is therefore much larger
than the splitting between the WIMP excited and ground states. 
Secondly, at the late stages of their evolution, neutron
stars can be rather cold objects due to lack of possible burning or
heating mechanisms, and therefore heating by annihilation of the dark
matter could produce an observable effect.

Close cousins of the neutron stars are the white dwarfs, the second
most compact objects. They are easier to observe due to their larger
surface area. However, they are lighter and less dense than neutron
stars. For an efficient capture, a dark matter particle has to collide
at least once per star crossing. For a neutron star, this requires the
cross section to satisfy $\sigma_N \gtrsim 10^{-45}\text{cm}^2$,
while for a solar mass white dwarf of radius $5000$~km one should
have $\sigma_N \gtrsim 10^{-39}\text{cm}^2$. As a result, neutron
stars can probe much smaller values of the WIMP-to-nucleon
cross section.

In this paper we consider constraints on the dark matter parameters
that may arise from the neutron star cooling. This question has been
addressed previously~\cite{Kouvaris:2007ay}. Here we concentrate
specifically on the effect of the dark-matter-rich environments such
as the Galactic center or cores of the globular clusters, and on the
role of the neutron star progenitor.

In section II we review the accretion and annihilation rates of dark
matter WIMPs relevant for neutron stars. In section III we study how
the accretion of dark matter on the progenitor of a neutron star
can affect the accretion and annihilation rates of WIMPs in a
neutron star emerging after the collapse of its progenitor. In section
IV, we present lower bounds for the surface temperature of a neutron
star as a function of its position in the galaxy, and in section V we
derive similar limits for neutron stars in globular clusters. We
conclude in section VI.

\section{Accretion and Annihilation of Dark Matter Inside a Neutron
  Star}

The accretion rate of the dark matter WIMPs onto a neutron star, having taken
into account relativistic effects~\cite{Kouvaris:2007ay}, is
\begin{equation}
F=\frac{8}{3} \pi^2 \frac{\rho_{\text{dm}}}{m} \left ( \frac{3}{2 \pi
  v^2} \right )^{3/2} \frac{GMR}{1-\frac{2GM}{R}}
v^2(1-e^{-3E_0/v^2})f, \label{accretion}
\end{equation}
where $\rho_{\text{dm}}$ is the dark matter density at the neutron
star location, $m$ is the mass of the WIMP, $M$ and $R$ are the mass
and the radius of the neutron star respectively, $v$ is the average
velocity of WIMPs asymptotically far from the star, and $E_0$ is the
typical WIMP energy loss at a single collision inside the star. The
energy $E_0$ defines the maximum energy of WIMPs that can be trapped
gravitationally after a single collision. The factor $f$ equals one
for both elastic and inelastic cross section if the latter is larger
than $\sim 10^{-45}~\text{cm}^2$, and equals
$f=\sigma_N/(10^{-45}~\text{cm}^2)$ for
$\sigma_N<10^{-45}~\text{cm}^2$. This factor describes inefficiency of
DM trapping in case the probability of collision during a single
passage through the star is less than one.  For a typical neutron star
of mass $M=1.4M_\odot$ and radius $R=10$~km, the rate of accretion is
\begin{equation}	
F=1.25 \times 10^{24} {\rm s}^{-1}
\left (\frac{\rho_{\text{dm}}}{\text{GeV}/\text{cm}^3} \right ) \left
(\frac{100 \text{GeV}}{m} \right )f, \label{rate}
\end{equation}
where we have used $v=220$ km/s.

We are interested in dark matter candidates that can self-annihilate.
In that case the number of dark matter particles inside the star
$N(t)$, as a function of time, is governed by
\begin{equation}
\frac{dN(t)}{dt}=F-C_AN(t)^2,
\label{ratediff}
\end{equation}
where $C_A=\langle \sigma_A v \rangle/V$ is the thermally averaged
annihilation cross section over the effective volume within which the
annihilation takes place. The solution of this equation is
\beq
N(t)= \sqrt{\frac{F}{C_A}} \text{Tanh} \frac{t+c}{\tau},
\eeq
where $c$ is a constant determined by the initial condition, and
\[
\tau=1/\sqrt{FC_A}.
\]
The power released inside the star
due to the annihilation of dark matter is
\begin{equation}
W(t)=Fm\,\text{Tanh}^2\frac{t+c}{\tau}.
\label{energy}
\end{equation}
Ignoring the initial constant $c$, the $\text{Tanh}$ saturates to one
for times larger than $\tau$, and the released power equals the
accretion rate times the dark matter mass, $W = Fm$.

The WIMP particles trapped in the neutron star further collide
with the neutrons and eventually come to thermal equilibrium with the
star. Pauli blocking plays important role in this process. Initially,
the WIMP velocity is large, so that the energy transferred to neutrons
exceeds the Fermi energy and all neutrons participate in the
scattering. Once the WIMP velocity drops and the transfer energy
becomes smaller than the Fermi energy, only a fraction of neutrons
close to the Fermi surface remains available for collisions. With the
account of the blocking, the time necessary for a WIMP to cool down to
a temperature $T$ is estimated as follows:
\[
t_{\rm th} = 0.2 {\rm yr} \left({m\over {\rm TeV}} \right)^2
\left({\sigma\over 10^{-43}{\rm cm}^2}\right)^{-1}
\left({T\over 10^5{\rm K}}\right)^{-1}.
\]
This estimate is in agreement with Ref.~\cite{Goldman:1989nd}.  For the
cross sections of order of the present experimental limit, $t_{\rm
  th}$ is a very short time compared to a typical age of neutron stars
(of order several Myr) that we consider here. For a typical WIMP of
1~TeV, the thermalization cannot be achieved in $1$~Myr if the cross
section is smaller than $10^{-51}$ $\text{cm}^2$. This bound might be
even lower considering that thermalization might have been achieved at
higher temperatures at a slightly earlier time than $10^6$ years.

If the thermalization condition is satisfied, the WIMPs follow a
Boltzmann distribution in velocities and distance from the center of
the the star, with almost all of the dark matter concentrated within
the thermal radius
\begin{equation}
r_{\text{th}}=\left (\frac{9T}{8\pi G \rho_c m} \right )^{1/2}\simeq
22 \text{cm} \left (\frac{T}{10^5 \text{K}} \right )^{1/2} \left (\frac{100
  \text{GeV}}{m} \right )^{1/2} ,\label{rth}
\end{equation}
where $\rho_c$ is the density of the neutron star.  If the
thermalization condition is not satisfied, the WIMPs occupy
a larger volume up to the total volume of the neutron star.

Let us now review the basics of neutron star cooling.  We will assume
here the usual picture of a neutron star, ignoring exotic
possibilities such as quark matter cores, color superconducting
matter~\cite{Alford:1997zt} and other effects that can alter the cooling of
a typical star~\cite{Alford:2004zr}. In the standard picture, the
neutron star cools through modified Urca process the first million
years, and later through thermal photon emission from its surface. As
it was pointed out in~\cite{Kouvaris:2007ay}, heating from dark matter
annihilation can compete with the photon emission once the temperature
of the star drops at later times. This happens when the power released
in dark matter annihilations, Eq.~(\ref{energy}), equals the thermal
energy loss rate, $L_{\gamma}=4\pi R^2 \sigma T^4$, where $\sigma$ is
the Stefan-Boltzmann constant, $R$ is the radius of the neutron star
and $T$ is the surface temperature of the star. For a given local dark
matter density, the photon emission and heating due to dark matter
annihilation equilibrate at a surface temperature
\begin{equation}
\frac{T}{10^5 \text{K}}=0.04 \left (
\frac{\rho_{\text{dm}}}{\text{GeV}/\text{cm}^3} \right
)^{1/4}. \label{tempe}
\end{equation}
Once the equilibrium is reached, the temperature does not drop further
but remains constant~\cite{Kouvaris:2007ay}.

The time needed to reach the equilibrium stage is determined by the
longest of the two time scales: the time of cooling of the neutron star to
sufficiently low temperature, and the time scale $\tau$ entering
Eq.~(\ref{energy}). The neutron star cooling takes at least 1~Myr
\cite{Kouvaris:2007ay}, while the value of $\tau$ is estimated as follows:
\begin{equation}
\tau=3.4 \times
10^{-5} \text{yr} \left (\frac{100}{m} \right )^{1/4} \left
(\frac{\text{GeV}/\text{cm}^3}{\rho_{\text{dm}}} \right )^{1/2} \left
(\frac{10^{-36}\text{cm}^2}{\langle \sigma v \rangle} \right )^{1/2}
\left (\frac{T}{10^5\text{K}} \right )^{3/4}
f^{-1/2}, \label{timescale}
\end{equation}
where $\langle \sigma v\rangle$ is the thermally averaged annihilation cross section.

In dark matter models where WIMPs are produced thermally, the
annihilation cross section is of the order of
$10^{-36}\text{cm}^2$. Since we are interested in times larger than a
million years, we see that for a typical thermal relic WIMP, $\tau$ is
by many orders of magnitude smaller than the neutron star cooling
time, and therefore Eq.~(\ref{tempe}) becomes applicable as soon as the
neutron star temperature drops at a sufficiently small value.

In models with non-thermal dark matter production, the annihilation
cross section is essentially a free parameter.
\begin{figure}[!tbp]
\begin{center}
\includegraphics[width=0.7\linewidth]{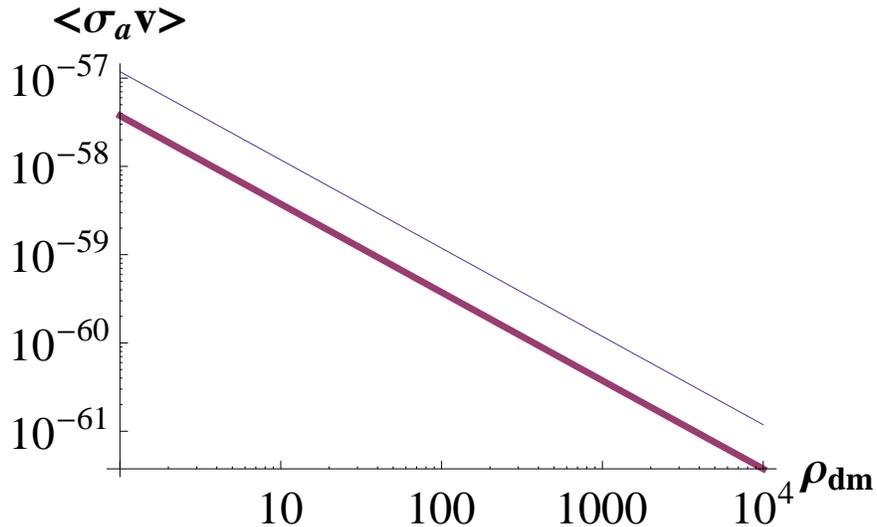}
\caption{The minimum annihilation cross section (in $\text{cm}^2$) as
  a function of the local dark matter density (in
  $\text{GeV}/\text{cm}^3$) for a dark matter WIMP of 100 GeV (thin
  line), and 1 TeV (thick line), in order the characteristic time
  scale for equilibrium between accretion and annihilation of WIMPs
  $\tau$ to be one million years (having assumed a temperature of
  $10^5$ K).}
\end{center}
\end{figure}
In Fig.~1 we plotted the minimum annihilation cross section required
to reach equilibrium between the dark matter accretion and
annihilation in 1~Myr, as a function of the local dark matter density
for two distinct values of the WIMP mass. The equilibrium neutron star
temperature is taken to be $10^5$ K. As it can be seen from the plot,
the minimum annihilation cross section ranges from $10^{-61}$ to
$10^{-57}$ $\text{cm}^2$. Therefore, the constraints on the dark
matter related to the dark matter annihilation in neutron stars are
valid even for non-thermally produced WIMPs with an extremely small
annihilation cross section. As we argue later, the accumulation of
dark matter in the neutron star progenitor extends the range of
applicability of the equilibrium equations to even lower annihilation
cross sections.

To conclude this section, let us comment on the WIMP accretion in
white dwarfs, another compact object where the dark matter
annihilation may lead to observable effects \cite{McCullough:2010ai}.
White dwarfs have typical radii two orders of magnitude larger than
neutron stars, which makes their observation easier.  However, being
less dense, white dwarfs lose significantly in the accretion rate. The
efficient capture of WIMPs in white dwarfs (i.e., at least one
collision per star crossing) requires a cross section larger than
$\sim 10^{-39}$~cm$^2$, while the corresponding number for a neutron star
is $\sim 10^{-45}$~cm$^2$. This is an important difference, in view of the
existing experimental limits. One might try to improve the situation
by assuming that the scattering of WIMPs off the whole nuclei of
matter constituting the white dwarf is coherent. In the case of carbon
nuclei, this would increase the probability of scattering by a factor
of 12.  However, taking into account the acceleration of the WIMPs due to the
gravitational field of the white dwarf, a typical WIMP of 100 GeV or
heavier has a de Broglie wavelength much smaller than the radius of
the carbon nucleus. Since the loss of coherence is
exponential~\cite{Lewin:1995rx}, the enhancement factor practically
disappears.  In addition, in case of Majorana fermions, WIMPs can only
have spin-dependent interactions with nuclei, so the interaction with
the whole $^{12}$C nucleus is zero.

\section{The effect of a neutron star progenitor}

Neutron stars are occasionally formed when a supermassive star
collapses gravitationally after having burned all of its fuel. The
collapse is followed by a supernova explosion (type II) and a
proto-neutron star is formed evolving eventually to a neutron star. In
this section we investigate what is the effect of the presence of the
massive star on the accretion and annihilation rates derived in the
previous section. In principle, the existence of a supermassive star
can change the local dark matter density by accretion, and therefore
once the neutron star is formed, it might accrete with a different
rate because of the different dark matter density. However, we will
see that this effect could have observational consequences only in
scenarios where WIMPs have extremely small annihilation cross
sections.

We start by considering a typical massive star of 15 solar masses. In
the first stage the star burns hydrogen for about 11 million years. It
is followed by the helium burning stage which lasts for about 2
million years, then by the carbon stage for 2000 years, and then by
neon, oxygen, and silicon stages before the star
explodes~\cite{Woosley:2002zz}. The last stages are very short and
have no significant effect on the accretion of the dark matter. In
fact, even the carbon stage can be neglected as the amount of dark
matter accreted at this stage is at least one order of magnitude
smaller than during the first two stages.

We estimate the amount of dark matter accumulated in the
progenitor star at the hydrogen and helium stages by means of
Eq.~(\ref{accretion}), where the relativistic effects can be neglected
and the fraction $f$ representing the probability of a collision
during single star crossing is now much smaller than one.
The total amount of the dark matter accumulated during the star
lifetime is
\begin{equation}
N_0 = 5\times 10^{37} \left({\rho_{\rm dm}\over {\rm GeV/cm}^3}\right)
\left({100~{\rm GeV}\over m}\right)
\left({\sigma_N\over 10^{-43}~{\rm cm}^2}\right).
\label{eq:DM-progen-total}
\end{equation}
Here, as before, $\rho_{\rm dm}$ is the dark matter density at the
star location. The helium state dominates the accretion, unless the
cross section is spin-dependent, in which case only the hydrogen stage
contributes. The total amount of the accumulated dark matter is then
reduced by roughly a factor of 5.

A remark is in order. From Eq.~(\ref{rate}) we find that the total
amount of WIMPs accreted by a neutron star in the characteristic time
of $1$~Myr is $4\times 10^{37}$, which is comparable to the total
amount of WIMPs accumulated by the neutron star progenitor. Thus, the
effect of the progenitor cannot be neglected a priori and has to be
investigated in more detail.

One may be surprised that two quite different objects --- the neutron
star progenitor and the neutron star itself --- accrete dark matter
with comparable efficiency. This numerical coincidence is a result of
two competing effects which roughly compensate each other. Larger mass
and radius of a progenitor lead to more dark matter particles crossing
the star. However, smaller matter density makes the probability of
capture after a single crossing small.

Now we turn to the question of how fast the WIMPs can be thermalized
inside the massive star. The thermalization determines the space
distribution of WIMPs at the moment of the neutron star formation.  We
estimated the thermalization time as the time it takes a WIMP to lose
kinetic energy down to the temperature of the star core.  At the
hydrogen stage, the WIMPs of $m=100$~GeV have enough time to
thermalize, while the WIMPs of $m=1$~TeV do not. At the helium and
subsequent stages, the thermalization is achieved for the whole
range of masses from $100$~GeV to $1$~TeV.

Once thermalized, the WIMPs follow the Boltzmann distribution in
velocities and distances from the center of the star, with most of
them concentrated within the thermal radius (\ref{rth}), where now $T$
is the core temperature and $\rho_c$ is the mass density at the
corresponding stage. Therefore, at the end of the star's life, most of
WIMPs will reside within the thermal radius of the last, silicon
stage, with $T=3.34 \times 10^9$ K and $r_{\rm th} = 2\times 10^7~{\rm
  cm}$. We also found that no significant amount of accumulated dark
matter annihilates during the lifetime of the massive star for
self-annihilation cross sections of the order of $10^{-36}\text{cm}^2$
or smaller.

Since the probability of WIMP scattering off a nucleon during single
star crossing is much smaller than one even at the silicone stage, the
supernova explosion will have no direct impact on the WIMP
distribution after the explosion. Therefore the newly-born neutron
star will find itself inside a dense dark matter cloud with a total
amount of dark matter given by Eq.~(\ref{eq:DM-progen-total}) and the
temperature and radius characteristic of the dark matter distribution
at the silicone stage.

When the neutron star is formed, it starts accreting the surrounding
dark matter, both from the cloud accumulated by the neutron star
progenitor and from the background dark matter distribution. Since the
dark matter is essentially non-interacting, these processes are
independent. The amount of WIMPS in the cloud, $N_c$, is governed by
the equation
\begin{equation}
\frac{dN_c}{dt}=-F= -1.25\times 10^{26} {\rm s}^{-1}
\left({{\rm cm}^3\over V}\right) f \times N_c,
\label{rate2}
\end{equation}
where $F$ is the rate given by Eq.~(\ref{rate}). The dark matter
density in the cloud drops exponentially with the characteristic time
scale
\[
\tau_{\rm acc} = 2\times 10^{-4} {\rm s}
\left({m\over 100{\rm GeV}}\right)^{-3/2}.
\]
Here we have assumed a spin-independent cross section and the volume
of the cloud that corresponds to the thermal dark matter distribution
at the silicone stage. The constant $f$ is set to 1. However, even for
a spin-dependent cross section, for which the dark matter cloud
is much larger as it corresponds to the hydrogen stage, the timescale
$\tau_{\rm acc}$ is of the order of a year. In either case,
independently of the initial value, the WIMP density in the cloud
drops down to the background density in a very short time, i.e., all of
the dark matter in the cloud gets accreted by the neutron star fast.

\begin{figure}[!tbp]
\begin{center}
\includegraphics[width=0.7\linewidth]{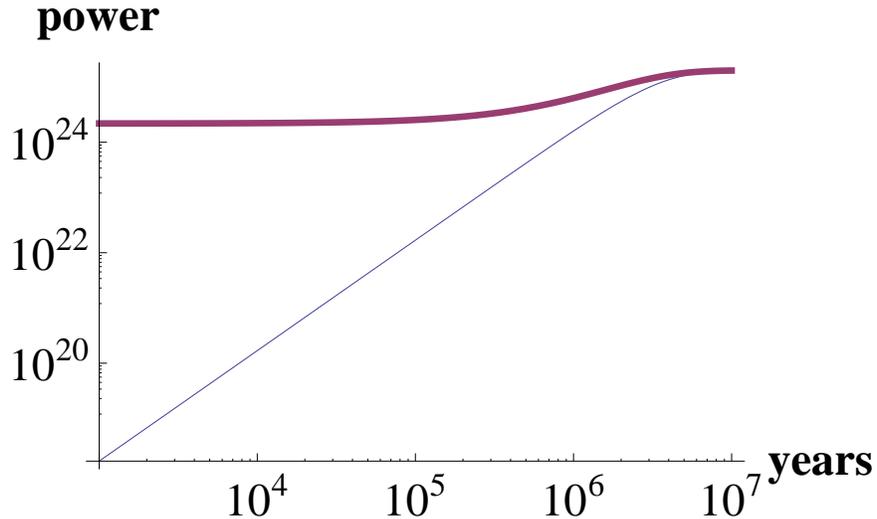}
\caption{Power due to WIMP burning in erg/sec as a function of the
  time. The thick line assumes the pre-existence of a massive star,
  where in the thin one, the neutron star starts accreting at $t=0$. We
  assumed an annihilation cross section of $10^{-60}\text{cm}^2$ and
  $\rho_{\text{dm}}=100$ GeV$/\text{cm}^3$.}
\end{center}
\end{figure}
Whether the accumulated dark matter produces an observable effect
depends on the annihilation cross section. From Eq.~(\ref{timescale})
we see that for annihilation cross sections typical of thermal relic
models, the characteristic time scale for reaching the equilibrium
between accretion and annihilation is very small. This means that very
shortly after the neutron star's birth, the dark matter accumulated by
the neutron star progenitor will be burned out. At the initial stages
of the neutron star cooling, this extra energy release is negligible in
the total energy balance determined by the modified Urca processes.

However, for non-thermal WIMP candidates with a very small cross
section, the time scale for reaching the equilibrium between accretion
and annihilation $\tau$, could be of the order of a million years. In
this case the dark matter accumulated by the neutron star progenitor
might have an observable effect as it speeds up the transition to the
equilibrium regime. In Fig. 2 we plotted the power produced by WIMP
burning as a function of the time. We assumed a very small
annihilation cross section $\sigma_A=10^{-60}\text{cm}^2$ and the background
dark matter density $\rho_{\text{dm}}=100$ GeV$/\text{cm}^3$ (see
Sects.~\ref{sec:neutron-stars-close} and
\ref{sec:neutr-stars-glob}). The difference is large at small times,
and decreases with time until both curves converge to the same
value. At $1$~Myr, the increase in the power due to the accumulated
dark matter is given by a factor of 4, corresponding into a temperature
increase by a factor of $\sqrt{2}$. Interestingly, in this particular
scenario the temperature continues to increase up to a neutron star
age of about $10^7$~yr.

Finally, consider the effect of the neutron star kick velocities.  In
principle, a kick given to a neutron star at collapse could move it
outside of the dark matter cloud accreted by the neutron star
progenitor. This would happen if in the rest frame of the neutron
star, the kinetic energy of an average WIMP after the kick, is larger
than its potential energy. Assuming that the accreted dark matter is
concentrated within the thermal radius of the silicon stage (roughly
$5000$ Km), it would take a kick of at least $8000$~km/s in order for
the WIMPs to escape from the gravitational well of the neutron
star. Since usually kicks are of the order of $1000$~km/s or less,
this effect cannot change our conclusions.

\section{Neutron Stars close to the Galactic Center}
\label{sec:neutron-stars-close}

\begin{figure}[!tbp]
\begin{center}
\includegraphics[width=0.7\linewidth]{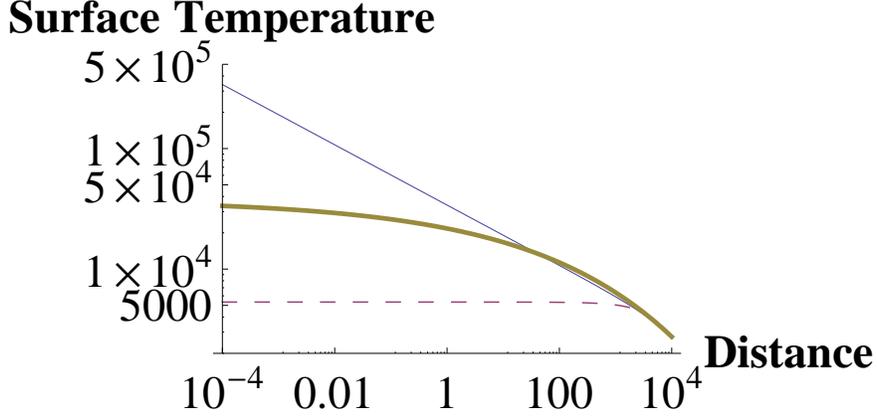}
\caption{The surface temperature of a typical old neutron star in
  units of K as a function of the distance of the star from the
  galactic center in pc, with the dark matter annihilation taken into
  account. The three curves correspond to three different dark matter
  profiles: NFW (thin solid line), Einasto (thick solid line), and
  Burkert (dashed line).}
\end{center}
\end{figure}
Eq.~(\ref{tempe}) gives the surface temperature of a neutron star that
can be sustained at later ages by burning dark matter. Obviously, the
effect is more pronounced in places where the local dark matter
density is high, such as the galactic center. The dark matter halo
profile of the Milky Way is the subject of intense
research. Several simulations have been performed in order to obtain
the profile that describes rotation curves best. We will
consider three different profiles. The first one is the standard
Navarro-Frenk-White (NFW)
\begin{equation}
\rho_{\text{NFW}}=\frac{\rho_s}{\frac{r}{r_s}(1+\frac{r}{r_s})^2},
\label{nfw}
\end{equation}
where $\rho_s=0.26$ GeV$/\text{cm}^3$, and $r_s=20$
kpc~\cite{Navarro:1995iw}. The second is the Einasto profile
\begin{equation}
\rho_{\text{Ein}} =\rho_s \text{exp} \left [-\frac{2}{\alpha} \left [
    \left (\frac{r}{r_s} \right )^{\alpha}-1 \right ] \right ],
\end{equation}
where $\rho_s=0.06$ GeV$/\text{cm}^3$, $\alpha = 0.17$, and $r_s=20$
kpc~\cite{Navarro:2008kc}. The third profile is the Burkert
\begin{equation}
\rho_{\text{Bur}}=\frac{\rho_s}{ \left (1+\frac{r}{r_s} \right ) \left
  [1+\left (\frac{r}{r_s} \right )^2 \right ]},
\end{equation}
where $\rho_s=3.15$ GeV$/\text{cm}^3$, and $r_s=5$
kpc~\cite{Burkert:1995yz}. All the profiles are normalized
to a local dark matter density at the Earth location of $0.3$~
GeV$/\text{cm}^3$.

In Fig.~3, we plotted the surface temperature of a neutron star that
can be maintained by burning accreted dark matter WIMPs for the three
aforementioned dark matter halo profiles. As expected, the Burkert
profile gives the smallest surface temperature almost regardless the
distance to the Galactic center. There is no spike in the Burkert
profile, and the local dark matter density increases only by one order
of magnitude when going from the location of the Earth to the galactic
center, making the neutron star surface temperature (that scales as
the local dark matter density to the power $1/4$) too small (around
5000 K) to be detected with the current observational
capabilities. The Einasto profile which is considered the best fit to
the data, gives a temperature slightly over $3 \times 10^4$ K, varying
slowly as a function of the distance to the galactic center. For
example, the surface temperature changes only by a factor of five as
one goes from the inner 1~kpc down to the inner $10^{-4}$~pc of the
galaxy. The NFW profile gives the most pronounced effect which is due
to the spike in the galactic center.  The surface temperature rises
faster (compared to the other profiles), reaching a value of $3 \times
10^5$ K at $10^{-4}$ pc from the Galactic center.

The curves of Fig.~3 should be considered as lower bounds for the
surface temperature of the neutron star: for a given dark matter halo
profile, the curves represent the coldest temperature the neutron star
can reach if it burns dark matter. The observation of a neutron star
with lower temperature would mean that dark matter WIMPs with an
elastic cross section as low as $10^{-45}\text{cm}^3$ (or even lower,
see below) and an annihilation cross section as small as
$10^{-57}-10^{-61}\text{cm}^2$ depending on the mass of the WIMP, and the local dark matter density, are
excluded. The observation of the neutron star with a surface
temperature higher than the predicted curves is consistent with the
existence of dark matter WIMPs with the above
characteristics. However, a mere existence of such a neutron star is not
conclusive for the existence of these WIMPs. A neutron star can have a
larger temperature because it might still be young, for example at the
stage where it still cools through neutrino emission via Urca
process, or even if equilibrium between photon emission and WIMP
burning has not yet been reached. In addition, non-isolated neutron
stars can maintain a relatively high temperature due to accretion of
ordinary matter from a binary companion star. In that case a
conclusive answer would require the observation of an isolated neutron
star and an accurate knowledge of its age.

As we already have mentioned, for an elastic (or inelastic) cross
section larger than $10^{-45}\text{cm}^2$, the mean free path of the
WIMP is smaller than the radius of the neutron star, ensuring that the
WIMP will collide at least once every time it passes through the
star. However, one can always trade the cross section for the local
dark matter density. In Fig.~4, we present the asymptotic surface
temperature of a typical star in the inner $10^{-4}$ pc of the galaxy
as a function of the elastic (or inelastic) cross section of the WIMP
with the neutron, for the NFW and Einasto profiles. As it can be seen
from the figure, even for a cross section as low as
$10^{-50}\text{cm}^2$, an old neutron star at $10^{-4}$ pc from the
galactic center may maintain a surface temperature of $2 \times
10^4$~K (using the NFW profile).

If WIMPs are fermions, the Pauli blocking may prevent the formation of
the black hole. The amount of dark matter needed to overcome the Pauli
blocking can be estimated by requiring that the gravitational
potential energy of a WIMP is larger than its Fermi energy,
\begin{equation}
\frac{GNm^2}{r}>k_F,
\end{equation}
where $N$ is the total number of WIMPs and $k_F=(3\pi^2 N/(4\pi
r^3/3))^{1/3}$ is the WIMP Fermi momentum. Taking into account
corrections due to the non-zero mass of the WIMPs and assuming that no
annihilations occur during the accretion, the number of WIMPs required
for a gravitational collapse is $10^{51}$ ($10^{48}$) particles for a
WIMP mass of 100~GeV (1~TeV). With an accretion rate of
Eq.~(\ref{rate}), even for an extremely high dark matter density
$\rho_{\text{dm}}=10^{10}$ GeV$/\text{cm}^3$ it would take $2.5\times
10^9$~yr ($2.5\times 10^7$~yr) in order for the WIMPs to collapse
gravitationally. If the dark matter density is smaller than
$10^{7}$~GeV/cm$^3$, the required number of WIMPs cannot be accreted
during the whole lifetime of the universe.  We should emphasize that
this is a conservative estimate, since we completely neglected the
annihilation during accretion. Therefore, the black hole formation by
dark matter inside neutron stars should not be of any concern in case
of fermionic dark matter except maybe in the very center of the
galaxy.
\begin{figure}[!tbp]
\begin{center}
\includegraphics[width=0.7\linewidth]{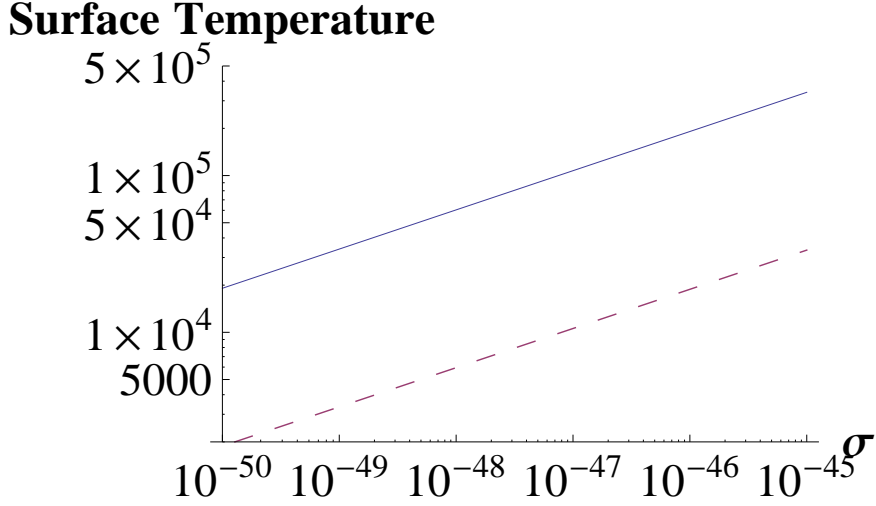}
\caption{The surface temperature of an old typical neutron star
  located at $10^{-4}$ pc in units of K as a function of the elastic
  (or inelastic) cross section (in $\text{cm}^2$) for the NFW (solid
  line), and the Einasto (dashed line) profiles.}
\end{center}
\end{figure}

\section{Neutron Stars in Globular Clusters}
\label{sec:neutr-stars-glob}

Another potentially dark-matter-rich environment is centers of
globular clusters. Globular clusters are dense spherical collections
of stars orbiting the galactic core.  A typical globular cluster, such
as M4, has a baryonic mass of $10^5$ solar masses and a core radius of
$0.5$ pc. Although globular clusters are baryon-dominated systems, the
dark matter density in their cores may exceed the average halo density
by several orders of magnitude.

To put constraints on the dark matter properties, we are interested in
old neutron stars with temperatures of order $10^5$~K or below. An
observation of an isolated neutron star with a temperature higher than
predicted by the cooling models may point, in the absence of other
heating sources, toward a WIMP-powered heating mechanism.

Although there are several globular clusters observed, we shall focus
on M4 (we will comment on the other clusters in the end of this
section). As an upper bound, we adopt an estimate of the dark matter
density in the core of M4 obtained in Ref.~\cite{Bertone:2007ae},
where a NFW dark matter density profile of Eq.~(\ref{nfw}) was used
with $\rho_s=24$ GeV$/\text{cm}^3$ and $r_s=171$
pc~\cite{Bertone:2007ae}. Apart from the parameters of the dark matter
density profile, the WIMPs in M4 have a much smaller velocity
dispersion compared to that of the galaxy. We use a value of
$v=20~\text{km}/\text{s}$~\cite{Gnedin:2002un}, which is an order of
magnitude smaller than that for the galaxy. By inspection of
Eq.~(\ref{rate}), we see that for a typical neutron star, the
accretion rate scales as inverse of the velocity, and therefore
reduction of the velocity by an order of magnitude implies an increase
of the accretion rate by the same factor.
\begin{figure}[!tbp]
\begin{center}
\includegraphics[width=0.7\linewidth]{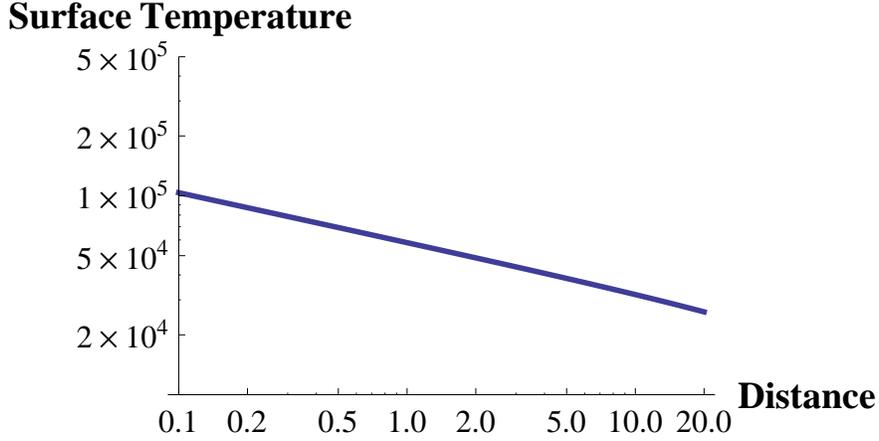}
\caption{The surface temperature of a typical old neutron star in
  units of K as a function of the distance in pc for a NFW profile of
  the globular cluster $M4$.}
\end{center}
\end{figure}

In Fig.~5 we present the surface temperature of an old neutron star
powered by the burning of dark matter for a NFW profile with the
parameters of the $M4$ globular cluster, as a function of the distance
from the center of the cluster. At 0.1 pc, the temperature is expected
to be $10^5$ K, while at the edge of the core (0.5 pc) the temperature
should be $7\times 10^4$ K.

Two comments are in order. First, the NFW profile used in the above
estimate does not include the effects of the tidal stripping and the
baryonic contraction. With these effects taken into account, the core
dark matter density in M4 is somewhat smaller. For example, close to
the edge of the core the difference is by about a factor of
3~\cite{McCullough:2010ai}, which corresponds to only to a $30\%$
decrease in the value of the surface temperature since the latter
scales as the dark matter density to the power $1/4$.

Second, although we considered $M4$ as an example, several globular
clusters have been observed that may contain such neutron stars. In
principle, the analysis should be redone for every particular
candidate. However, we should emphasize that $M4$ is a typical
globular cluster, and we do not expect to have dramatic changes in our
predictions. For example, 47~Tuc is another globular with a mass
larger than $M4$. The core dark matter density might be higher in
47~Tuc, but this would barely make a difference due to the slow
dependence of temperature on the dark matter density.

Although white dwarfs are easier to spot in globular clusters, the
neutron stars might be more appropriate for probing smaller
annihilation and elastic (inelastic) cross sections. Pulsars have
already been detected in globular clusters. A characteristic example
is the pulsar $1620-26$ found in the outskirts of the core of $M4$. It
is a part of a triple bound system with a planet and a white
dwarf. Since the age of $M4$ is of the order of billion years, the neutron
star is expected to be old enough to exhibit the effect of dark matter
burning. In this case its temperature should be $7\times 10^4$~K
(without baryonic contraction) or $5.3\times 10^4$~K (with baryonic
contraction).  However, the neutron star might still be accreting
matter from the companion white dwarf, in which case its temperature
may be higher than this prediction. Observation of the temperature
lower than $\sim5\times 10^4$~K would be inconsistent with the
existence of WIMP with the properties described above.

Another example is the source $X7$ of the 47~Tuc globular cluster
which has low variability and seems to be a non-accreting neutron
star, with a spectrum that fits a neutron star with a hydrogen
atmosphere and an effective temperature as low as
89~eV~\cite{Rybicki:2005id}. This is a temperature of the order of
$10^6$~K, an order of magnitude higher than predicted by the dark
matter burning mechanism. In the absence of the precise knowledge of
the neutron star age, this is consistent with the dark matter WIMPs
scenario. We should also mention that a detailed comparison requires a
more thorough investigation since there is no direct equivalence
between the Hydrogen atmosphere spectrum and the blackbody one. In
general, isolated neutron stars (or neutron stars that have low
variability and therefore no accretion from companion stars) in
globular clusters constitute ideal candidates for observing the
effects of the WIMP burning.

\section{Discussion}
In this paper we examined the effect of WIMP annihilation on the
temperature of a neutron star. We estimated the surface temperatures
of old neutron stars according to their location in the galaxy or in a
globular cluster. We also investigated the effect of a neutron star
progenitor on the accretion of WIMPs onto the neutron star.  We found
that, although a considerable number of WIMPs is accumulated by the
progenitor during the evolution preceeding the formation of a neutron
star, the effect of this accumulation is observable only in cases where
the annihilation cross section is extremely small.

We argued that observations of neutron stars with low (of order
$10^5$~K or lower) surface temperature will put constraints on a large
set of dark matter candidates.  Due to their high density, the
neutron stars accrete the dark matter at a significant rate even when
the WIMP-to-nucleon cross section (elastic or inelastic) is as low as
$10^{-45}\text{cm}^2$, which is two orders of magnitude lower than the
current experimental limit. Even for lower values of the cross
section, the effects of WIMP accretion and annihilation may be
observable in neutron stars which are situated in dark-matter-rich
environments such as the galactic center and cores of globular
clusters. Thus, the neutron stars can probe much smaller WIMP cross
sections than less dense objects such as, e.g., white dwarfs.

The WIMP constraints that we presented are valid even if the WIMPs
have a very small annihilation cross section as low as
$10^{-57}\text{cm}^2$ (or even lower for large local dark matter
densities). This means that our constraints hold also for a variety of
WIMP candidates that are produced non-thermally, for which the
annihilation cross section is, in  general,  a free parameter.

Perfect candidates to test the WIMP-burning heating mechanism are
isolated neutron stars (i.e., not showing accretion of ordinary matter
from other objects), which are old but appear warmer than predicted by
the conventional cooling models. There is a couple of examples of such
candidates. One of them is J0437-4715, a few billion years old neutron
star with a roughly $10^5$~K
temperature~\cite{Kargaltsev:2003eb}. Although this temperature can be
sustained by WIMP burning, it would require a substantial local dark
matter density, which is unlikely as J0437-4715 is only 140 pc from
the Earth. Unless there is a peak in the dark matter density at the
position of J0437-4715, WIMP burning cannot explain this
temperature. A similar candidate is J0108-1431 at 130~pc from Earth,
with a temperature $\sim 9 \times 10^4$ K~\cite{Mignani:2008jr}. Like
in the case of J0437-4715, this temperature is still higher than what
the dark matter burning can provide, assuming the dark matter density
at the location of J0108-1431 is the same as around the Earth.
Candidates like the above, with smaller temperatures or in rich dark
matter regions such as globular clusters, might make it possible to
constrain a large class of dark matter WIMP scenarios.

\section{Acknowledgements}
We would like to thank Sanjay Reddy, Dany Page, Konstantin Postnov and
David Kaplan for useful discussions. This work is supported in part by
IISN, Belgian Science Policy (under contract IAP V/27).

\end{document}